# Temperature dependence of the ohmic conductivity and activation energy of Pb$_{1+y}$(Zr$_{0.3}$Ti$_{0.7}$)O$_3$ thin films

D. P. Chu, B. M. McGregor, and P. Migliorato

*Epson Cambridge Laboratory, 8c King's Parade, Cambridge CB2 1SJ, UK*

*Department of Engineering, University of Cambridge, Cambridge CB2 1PZ, UK*

C. Durkan and M. E. Welland

*Department of Engineering, University of Cambridge, Cambridge CB2 1PZ, UK*

K. Hasegawa and T. Shimoda

*Seiko Epson Corporation, BTRC, Fujimi Plant, 281 Fujimi, Fujimi-machi, Suwa-gun, Nagano-ken, 392-8502, Japan*

The ohmic conductivity of the sol-gel derived Pb$_{1+y}$(Zr$_{0.3}$Ti$_{0.7}$)O$_3$ thin films (with the excess lead y=0.0 to 0.4) are investigated using low frequency small signal alternate current (AC) and direct current (DC) methods. Its temperature dependence shows two activation energies of 0.26 and 0.12 eV depending on temperature range and excess Pb levels. The former is associated with Pb$^{3+}$ acceptor centers, while the latter could be due to a different defect level yet to be identified.

The application of ferroelectric thin films in nonvolatile semiconductor memories[1,2] has stimulated interest in the study of their electrical properties. Lead zirconate titanate Pb(Zr,Ti)O$_3$ (PZT), as one of the two ferroelectric materials actually used in volume production of ferroelectric random access memories (FeRAMs), has been the subject of extensive studies. However, as it has been pointed out[3], the leakage current characteristics of the PZT thin films have received little attention compared to their ferroelectric properties. In view of requirement of reduced film thickness for high density FeRAMs such an investigation is clearly warranted.

The current-voltage (I-V) characteristics of a ferroelectric film is normally measured in the configuration of metal-insulator-metal (MIM). The leakage current at not very high electric field can usually be divided into regions in which one or more of the three conduction mechanisms dominates: ohmic, injection and space charge limited (SCL). The ohmic conduction is a result of the net current of thermal charge carriers inside the film. The injection conduction is due to the charge carriers getting over the Schottky barrier at the metal-insulator interface, while the SCL conduction occurs when the density of injected charge carriers overtakes that of the thermally generated. There have been studies of the injection and SCL conduction in PZT thin films[3-8] but ohmic conduction and in particular its temperature dependence have been largely ignored.

The amount of excess Pb in a PZT film shows considerable influence on the remnant polarization[8-10] and formation of a single phase material.[10,11] It also affects the crystalline phase[12] and high temperature (above T$_c$) conductance[13-15] in PZT ceramics. However, there has been little study on the effect of excess Pb to the room temperature conductance of a PZT film, which has a direct impact on the reliability and stability of FeRAM.

In this work, we analyze the ohmic conductivity as a function of temperature and composition for some PZT films of 200 nm thickness with different Pb compositions.

The PZT films were fabricated by sol-gel process, a method whereby the film composition can be easily changed. Five different levels of Pb in the sol-gel precursor were used to form the films of $Pb_{1+y}(Zr_{0.30}Ti_{0.70})O_3$ with the excess lead y=0.0, 0.1, 0.2, 0.3 and 0.4. The films were spin coated first, then dried at 180 C for 10 mins, followed by 400 C treatment for another 10 mins. Subsequently they were annealed in oxygen ambient at 600 C and 750 C for 5 mins and 1 min, respectively, resulting in polycrystalline PZT films 200 nm thick. Platinum and iridium were used, respectively, to form both the top and bottom electrodes. The films with Pt electrodes and y≥0.3 and those with Ir electrodes and y=0.4 showed little sign of polarization and very small resistance. Such a behavior suggests the presents of a large amount of metal precipitates. These samples were not considered in the present analysis.

Since ferroelectric films polarize in external electric field, we depolarized the films before each measurement in order to minimize the effect and allow meaningful comparisons of the results. This was done by applying an AC external field with continuously decreasing amplitude through a simple circuit which allows discharge of the film to be completed simultaneously.

It is well established that the power index n=1 and n≥2 in the expression I ~ $V^n$ indicate the ohmic and SLC conduction, respectively.[16,17] Fig. 1(a) shows a typical DC I-V curve (y=0.0 with Ir electrode). Two different conduction regions, the ohmic region with n=1.03 at V≤0.1 V (5 kV cm$^{-1}$) and SCL region with n=2.06 at around 0.8 V (40 kV cm$^{-1}$), are obvious. The positive voltage refers to the top electrode and the bottom electrode is grounded. Using this criterion, we were able to identify the DC ohmic conductivity of our samples. We also measured the complex impedance in the frequency range $10^{-2}$-$10^4$ Hz by using a small signal excitation (50 mV). The resonance associated with bulk capacitance enables to separate the bulk resistance from the contribution of the contact region, yielding an unambiguous determination of the bulk resistance.[18] Both the DC and AC ohmic conductivities, $\sigma_{ohmic}$, are shown in Fig. 1(b). We find a good agreement between the data for the positive, negative applied DC voltages and the AC measurements. This confirms that $\sigma_{ohmic}$ is the bulk conductivity of our films. The discrepancy between the AC and DC results for the highly resistive films with Pt electrodes may be due to the much longer relaxation time involved.

The temperature dependence of $\sigma_{ohmic}$ for Pt contacts with y=0.0 is shown in Fig. 2(a). Two different activation energies, $E_{a,LT}$ and $E_{a,HT}$, obtained by fitting the expression $\sigma_{ohmic}(T) = \sigma_0 \exp(-E_a/k_B T)$ to the measured data, are clearly identifiable at below and above 100 C, respectively. The activation energies as a function of excess Pb are plotted in Fig. 2(b). For y between 0.0 and 0.2 the films with Pt contacts have $E_{a,HT}$=0.2-0.3 eV and $E_{a,LT}$=0.09-0.12 eV. In films with Ir contacts, the activation energy is not affected by temperature but undergoes an abrupt decrease, from 0.26 eV to 0.12 eV, when the excess Pb level is increased above y=0.1. We also calculated the quantity of $N\mu = \sigma_0/e$, where $N$ is the density of charge carriers when $k_BT \gg E_a$ and $\mu$ the corresponding mobility. The results are shown in Fig. 3

The activation energy for Ir contacted samples with y≤0.1 and $E_{a,HT}$ of all Pt contacted samples coincide within the experimental errors. Similarly the activation energy for Ir contacted samples with y≥0.2 and $E_{a,LT}$ of all Pt contacted samples are also coincident within experimental errors. For simplicity we refer to these two energies as $E_1$ and $E_2$, respectively, with $E_1 < E_2$.

An acceptor level at ~0.26 eV has been reported for PZT thin films and associated with the $Pb^{3+}$ centers which are the $Pb^{2+}$ ions acting as hole traps[19-21]. We suggest that this center is responsible for the activation energy $E_2$. This implies[22] that the material is strongly compensated (p<<$N_A$). We attribute the higher $\sigma_{ohmic}$ in the films with Ir electrodes to a higher density of acceptor levels/lower density of compensating centers than present in the films with Pt electrodes. The origin of $E_1$ at around 0.12 eV is not clear. The value itself is about one half of 0.26 eV. However, the two activation energies cannot be associated with the same energy level via change in compensation ratio because of the sequence of $E_1$ and $E_2$ with temperature would be opposite to that shown in Fig. 2(a). We are left with the possibility of an unknown acceptor center or some other conduction mechanism.

Recently, the migration enthalpy of oxygen vacancies as low as 0.12 eV was reported in the PZT thin films,[23] in addition to the early measurement of the activation energy of 0.11 eV for PZT ceramics between 500 and 700 C[15]. Taking the room temperature mobility of oxygen vacancy as $10^{-8}$ cm$^2$ V$^{-1}$ s$^{-1}$,[24] its density will be no less than $10^{21}$ cm$^{-3}$ for the Ir electrode material with y=0.2 and 0.3, which is unrealistic. The work on the electrons trapped by dipolar defects consisting of oxygen vacancies and impurity Pt ions showed a binding energy of 0.1 eV,[25] but there was no mention of the case with impurity Ir ions. Besides, the large increment of excess lead levels in our samples makes it difficult for us to identify whether there is a transition from acceptor dominated state to donor dominated state. Another possibility could be electronic conduction by hopping. Hopping conduction is enhanced by structural or compositional disorders which may occur in high excess Pb samples. This could only be confirmed by extending the measurements to much lower temperatures.

In summary, we have developed a procedure to extract the bulk conductivity of ferroelectric PZT thin films and applied it to the analysis of the ohmic conductivity as a function of temperature and excess Pb concentration. Two regimes are observed characterized by activation energies of 0.26 and 0.12 eV, respectively. The former can be explained as the conductivity behavior of a strongly compensated p-type semiconductor with the Pb$^{3+}$ acceptor centers. The latter could be due either to an as yet unidentified acceptor center or hopping conduction. These two regimes are present in both Ir and Pt contacted samples, but the magnitudes of the associated conductivities vary which suggests that the concentration of acceptor centers or compensating centers is affected by the contact material.

The authors would like to thank Prof Richard Friend of Cavendish Laboratory, Cambridge University for helpful discussions.

**Figure captions**

Figure 1: (a) the DC I-V curve of the PZT film of y=0.0 with Ir electrodes; (b) the DC and AC $\sigma_{ohmic}$ of the PZT samples with different levels of excess lead, y, at room temperature. The +V and -V indicate the polarity of the DC voltage applied to the top electrodes and Pt and Ir are of the electrode materials, respectively.

Figure 2: (a) the temperature dependence of $\sigma_{ohmic}$ of the PZT sample with Pt electrodes and y=0.0; (b) the activation energies $E_a$ vs. level of excess Pb level y.

Figure 3: The product of charge density and mobility vs level of excess Pb. The subscript *i* refers to different corresponding activation energies.

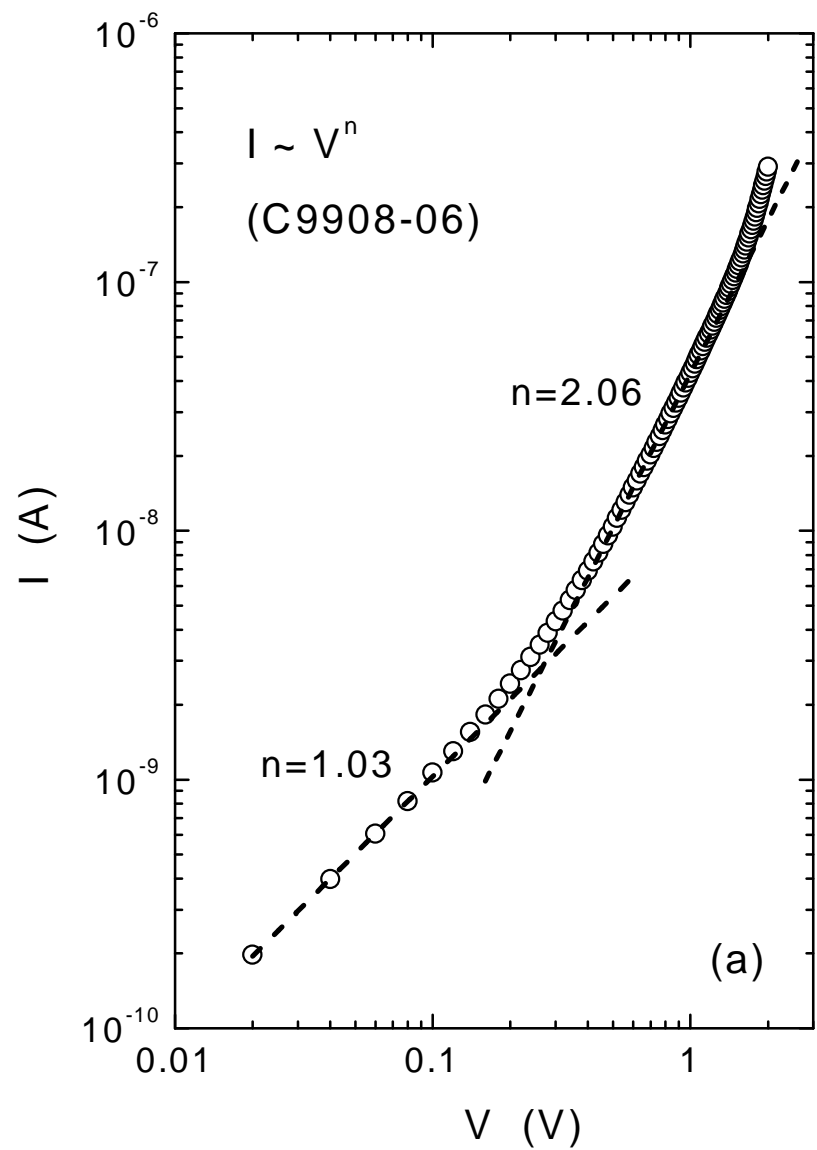
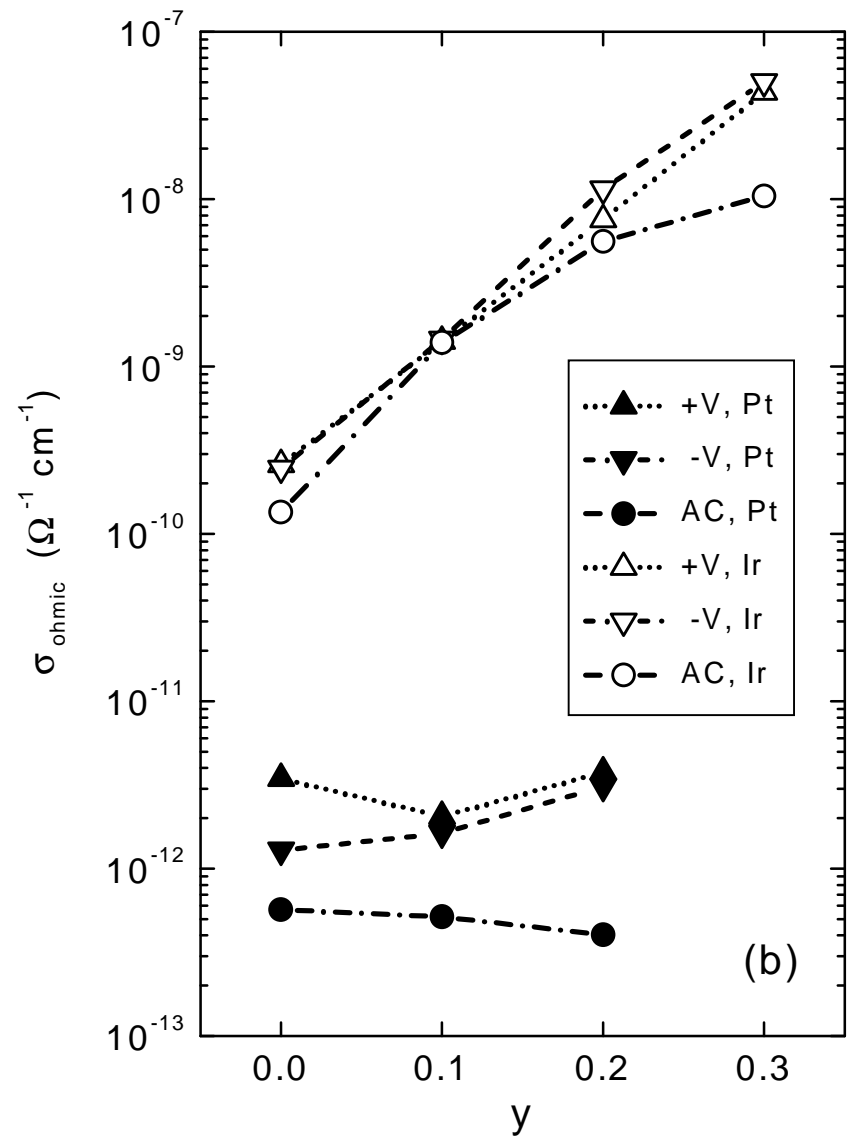

11 **Figure 1(a) and 1(b), D P Chu, et al.**

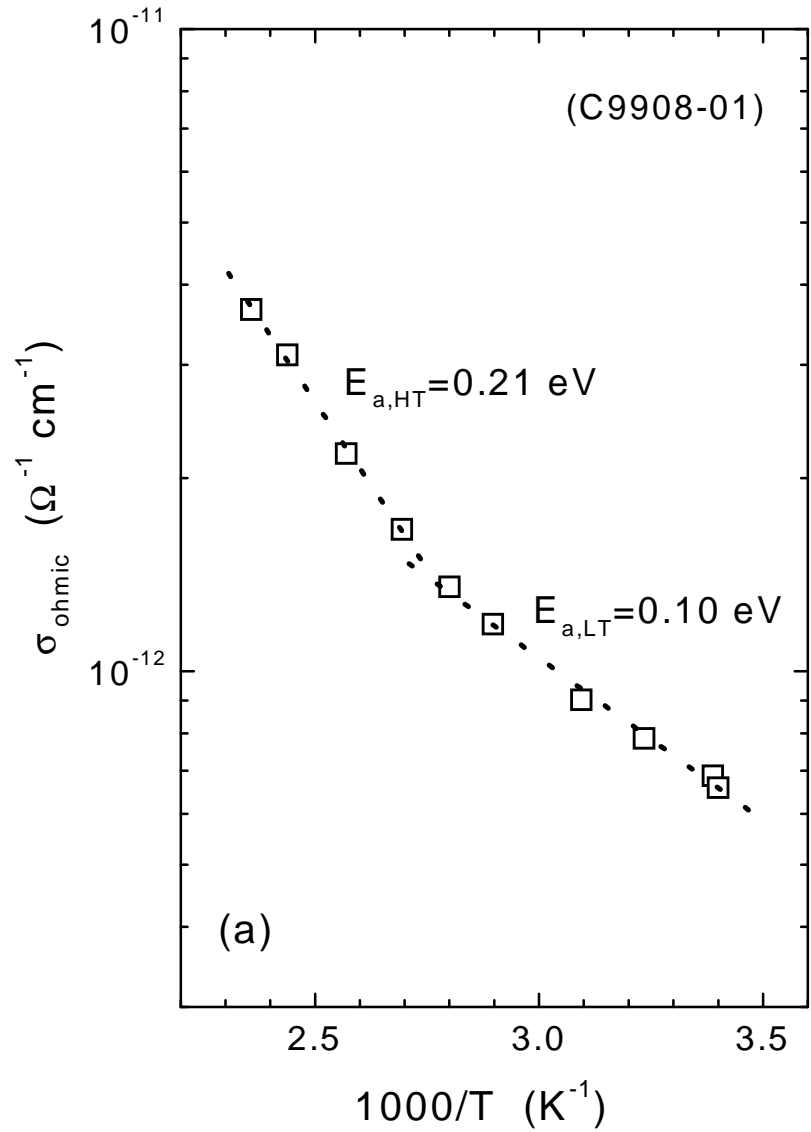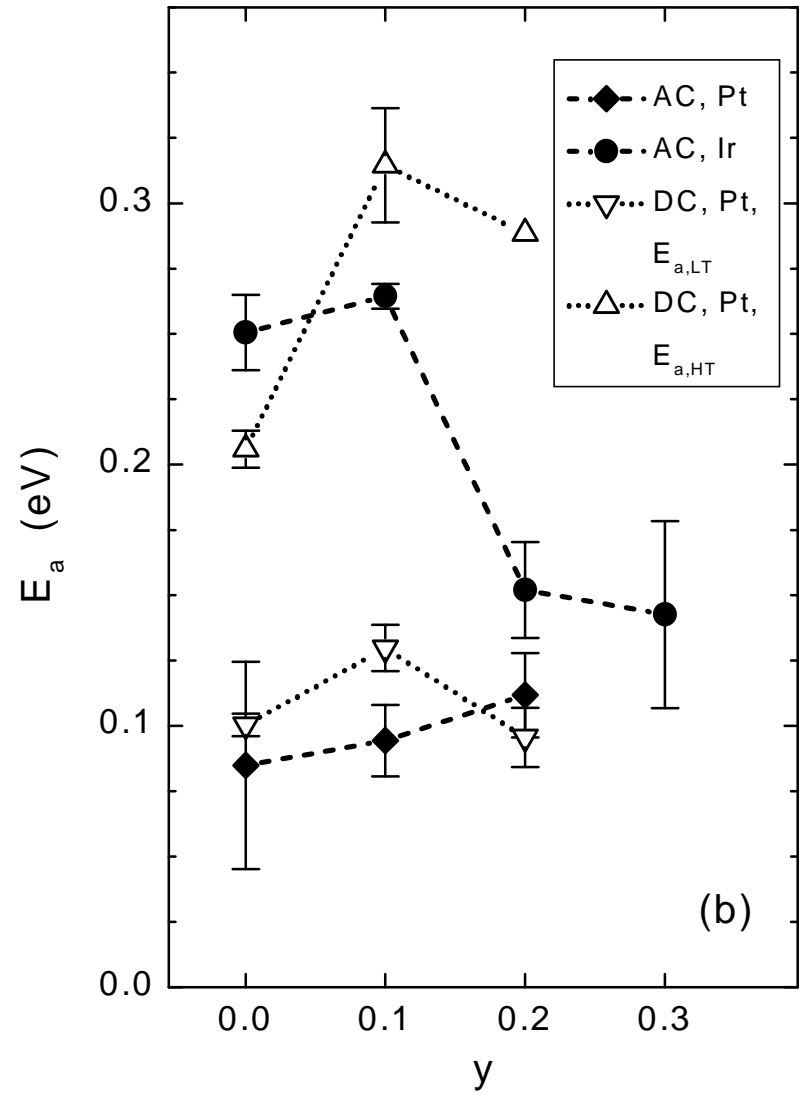

**Figure 2(a) and 2(b), D P Chu, et al.**

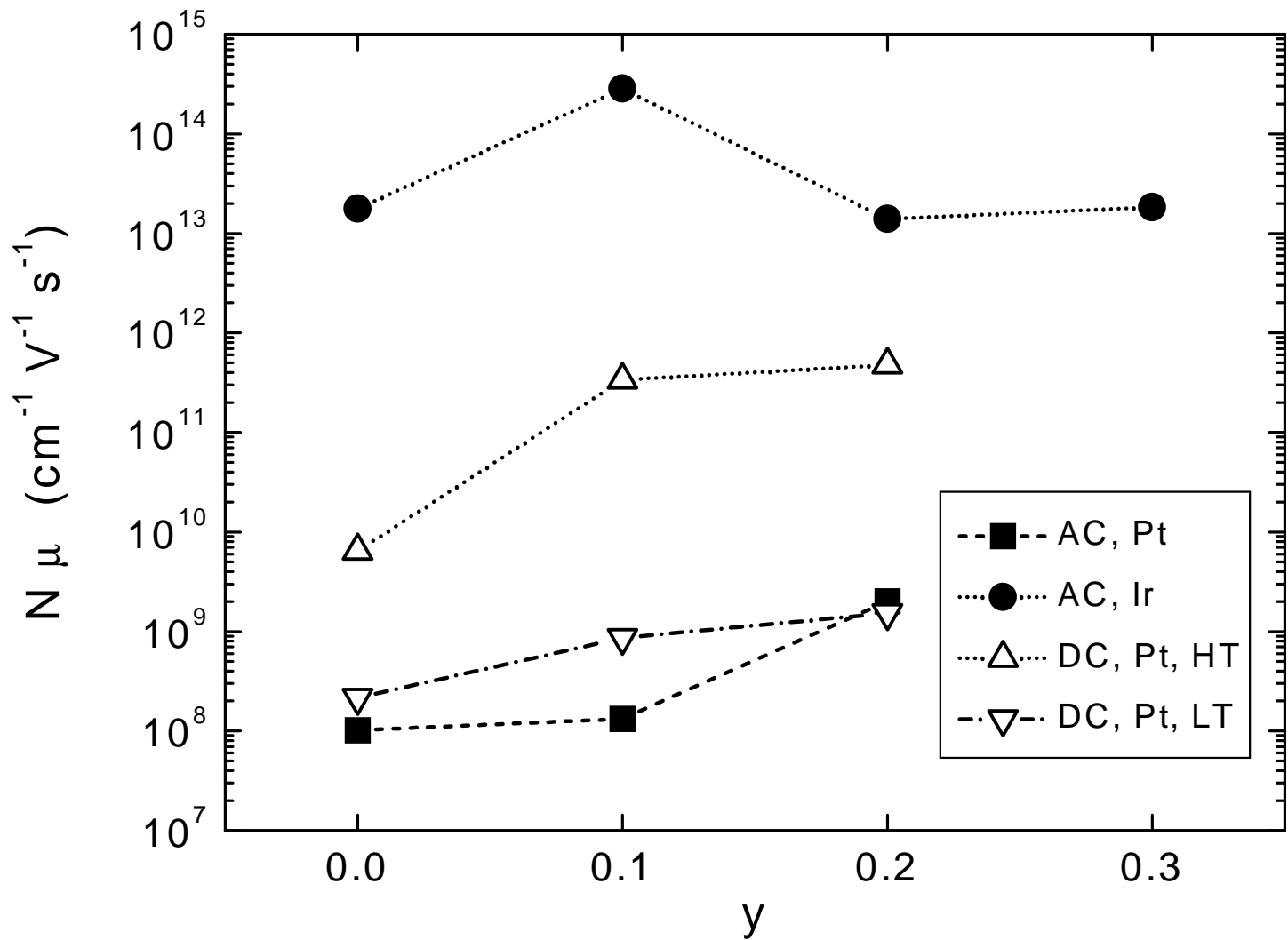

**Figure 3, D P Chu, et al.**